\documentclass[nofootinbib,twocolumn]{revtex4}
\usepackage{graphicx}
\usepackage{latexsym}
\def\be{\begin{equation}}
\def\ee{\end{equation}}
\def\bea{\begin{eqnarray}}
\def\eea{\end{eqnarray}}

\begin{document}

\title{Generating scale-invariant tensor perturbations in the non-inflationary universe}

\author{Mingzhe Li}
\email{limz@ustc.edu.cn}
\affiliation{Interdisciplinary Center for Theoretical Study, University of Science and Technology of China, Hefei, Anhui 230026, China}

%\date{\today.}

\begin{abstract}
It is believed that the recent detection of large tensor perturbations strongly favors the inflation scenario in the early universe. This common sense depends on the assumption that Einstein's general relativity is valid at the early universe. In this paper we show that nearly scale-invariant primordial tensor perturbations can be generated during a contracting phase before the radiation dominated epoch if the theory of gravity is modified by the scalar-tensor theory at that time. The scale-invariance protects the tensor perturbations from suppressing at large scales and they may have significant amplitudes to fit BICEP2's result.
We construct a model to achieve this purpose and show that the universe can bounce to the hot big bang after long time contraction, and at almost the same time the theory of gravity approaches to general relativity through stabilizing the scalar field. Theoretically, such models are dual to inflation models if we change to the frame in which the theory of gravity is general relativity. Dual models are related by the conformal transformations. With this study we reinforce the point that only the conformal invariant quantities such as the scalar and tensor perturbations are physical. How did the background evolve before the radiation time depends on the frame and has no physical meaning. It is impossible to distinguish different pictures by later time cosmological probes.
\end{abstract}
%\hfill USTC-ICTS-14-08
\maketitle

\hskip 1.6cm PACS number(s): 98.80.Cq, 98.80.Bp \vskip 0.4cm

Cosmic inflation \cite{inflation} is a hypothesis about the early universe.
It states that at the early time before the radiation dominated epoch (we call this time pre-big bang in this paper) our universe
experienced a period of nearly exponential expansions. This paradigm
is very successful. It provides not only solutions to the horizon and flatness puzzles existed in the
hot big bang cosmology but also mechanisms to generate initial density perturbations for structure formation.
The single field slow-roll inflation models generically predict adiabatic, nearly scale-invariant and Gaussian primordial density perturbations, which were confirmed with strong confidences by the observations of the cosmic microwave background radiation (CMB) \cite{Ade:2013ktc}.
Besides the scalar (density) perturbations, the single field inflation models also predict nearly scale-invariant and sizeable tensor perturbations, i.e., primordial gravitational waves, with the amplitude proportional to the energy scale at which the inflation taking place. The tensor perturbations left a unique imprint to observations by producing the curl-like B-mode polarizations on the CMB sky.
Recently the BICEP2 collaboration announced the detection of the B-mode polarizations of CMB at large angular scales \cite{Ade:2014xna}. This suggested a large tensor-to-scalar ratio, $r=0.20^{+0.07}_{-0.05}$ at $68\%~CL$ if all of these B-mode polarizations are originated from the primordial tensor modes. It is commonly believed that this result, if confirmed, strongly favors the single field inflation models. Some alternative models, such as the Ekpyrotic/Cyclic universe \cite{Ekpyrotic1}, bouncing universe \cite{Novello:2008ra} and so on usually predict non-detectable tensor modes and are disfavored by BICEP2's result.

This can be seen from the following general arguments. The single field model in the context of Einstein's gravity has the action $S=(1/2)\int d^4 x\sqrt{g}[R+\partial_{\mu}\phi\partial^{\mu}\phi-2V(\phi)]$, where we have used the unit in which $M_p=1/\sqrt{8\pi G}=1$. The tensor perturbation from $ds^2=a^2d\eta^2-a^2(\delta_{ij}+\gamma_{ij})dx^idx^j$, which is traceless and transverse $\gamma_{ii}=\partial_i \gamma_{ij}=0$, has the following quadratic action
\be\label{action1}
S^T=\frac{1}{8}\int d^4x a^2 [(\gamma_{ij}')^2-\partial_l\gamma_{ij}\partial_l\gamma_{ij}]~.
\ee
This is a massless spin-2 field coupled to the background through the cosmic scale factor $a$.
The prime means the derivative with respect to the conformal time $\eta$. The tensor perturbation has only two components denoted by $\lambda=+, \times$. Its re-scaled amplitude $v_{k}$, which relates to $\gamma_{ij}$ in Fourier space as
$\gamma_{ij}(\vec{k})=\sum_{\lambda}[v_{k}\hat{a}(\vec{k},\lambda)+v_{k}^{\ast}\hat{a}^{\dag}(-\vec{k},\lambda)]e_{ij}(\vec{k},\lambda)/a$, here $e_{ij}$ is the polarization tensor and $\hat{a}$ and $\hat{a}^{\dag}$ are the annihilating and creating operators of gravitons respectively, satisfies the equation of motion
\be
v_{k}''+(k^2-\frac{a''}{a})v_{k}=0~.
\ee
By quantization and choosing the Bunch-Davies vacuum at initial time $\eta\rightarrow -\infty, ~v_{k}= e^{-ik\eta}/\sqrt{2k}$, the resulted tensor power spectrum at later time when $\eta\rightarrow 0$ is $P_T=4k^3/(\pi^2)|v_k/a|^2\sim (-\eta)^{1-2\nu-4/(1+3w)}k^{3-2\nu}$. Here we have assumed the background equation of state $w$ is a constant and $\nu=|3(w-1)/2(1+3w)|$. We also for convenience put the time of all the pre-big bang phase at the range $-\infty <\eta<0$, so the time $\eta=0$ is approximately the beginning of the hot expansion. Taking the Ekpyrotic/Cyclic universe \cite{Ekpyrotic1} as the example, the perturbations of the cosmological scales were generated at the slow contracting phase in which $w>1$, hence the tensor spectral index $n_T=3-2\nu>2$. This spectrum has a large blue tilt and the power is suppressed deeply to undetectable level at large scales corresponding to the observation of BICEP2.
By contrast, the inflation, during which $w\simeq -1$ and $a\sim 1/(-\eta)$, predicts a constant and nearly scale-invariant tensor spectrum, i.e., $P_T=const.,~n_T\simeq 0$. So the power is not suppressed at large scales and can be significant. In addition, the contracting universe dominated by the matter in which $w\simeq 0$ \cite{Wands:1998yp} can also produce a nearly scale-invariant tensor spectrum. In this scenario the matter contraction must be interrupted by other phases well before bouncing to the hot big bang as indicated in the matter bounce models \cite{Cai:2013kja}, otherwise both the scalar and tensor perturbations, scaling as $(-\eta)^{-6}$ during the matter contraction, will blow up and make the background unstable. Another problem the matter contraction models encountered is that the classical anisotropies which scale as $a^{-6}$ are not suppressed. So it seems that the inflation scenario has the strong preference over other models if BICEP2's result is confirmed.

According to above observations, if the tensor spectrum is not strongly blue tilted, it is possible to have significant tensor perturbations at large scales. Currently there are some studies on the tensor spectral index, see e.g.,\cite{Cheng:2014bma}, but the confidence level is low and we need more data to improve it. But if we get a nearly scale-invariant tensor spectrum, as inflation predicted, we have the possibility to obtain significant tensor modes at large scales. In this paper we will focus on the production of the nearly scale-invariant tensor spectrum. The tensor perturbation couples to the background by the factor $a^2$ in the action (\ref{action1}). A few calculations show that the scale-invariant perturbation with constant amplitude can only be achieved if the background is nearly de Sitter, i.e., inflation where $a=-1/(H_I\eta)$ and the Hubble parameter $H_I$ a constant. This conclusion is made based on the assumption that the theory of gravity is Einstein's general relativity. If the theory of gravity is modified in the early universe, it is possible to obtain scale-invariant tensor perturbation without inflation. We can simply argue that if there are non-minimal couplings to the curvature scalar, $S=(1/2)\int d^4x \sqrt{g}[F(\phi) R+...]$, the quadratic action (\ref{action1}) will be modified as $S^T=(1/8)\int d^4x (a^2F) [(\gamma_{ij}')^2-\partial_l\gamma_{ij}\partial_l\gamma_{ij}]$. Now scale-invariance requires $a\sqrt{F}$ instead of $a$ scales as $ 1/(-\eta)$. That is to say even the spacetime deviates de Sitter significantly, we may still get scale-invariant tensor perturbation through the time dependence of the non-minimal coupling function $F$.

This possibility has been investigated in Refs. \cite{Piao,Qiu}, in which the authors constructed the models in the context of scalar-tensor theory to show that nearly scale-invariant scalar and tensor perturbations can be produced in the
slow expanding universe. In this paper we will pursue the productions of nearly scale-invariant perturbations during the contraction of the universe based on similar models.  Scalar-tensor theories are usually used to model the later time acceleration of the universe or the variation of the Newton constant \cite{Clifton:2011jh}. As we know, general relativity passed all the experiments at low energy scales. But at the early universe the energy scale is very high and the theory of gravity may get modified and replaced by a scalar-tensor theory. In fact scalar-tensor theories arise naturally from fundamental theories with higher dimensions. All versions of string theory predict the scalar-tensor theory rather than general relativity as the actual theory of gravity, in which the spin-2 graviton has a spin-0 partner, the dilaton. These theories are effective at high energy scales and at low scales they should approach general relativity since the solar system experiments have put stringent constraints on the deviations from general relativity. We also adapt this point here, at the early universe (pre-big bang) the theory of gravity is scalar-tensor and approaches general relativity at the post-big bang era.
The model we consider has the general action
\be\label{action}
S=\int d^4x \sqrt{g} [\frac{1}{2} F(\phi) R+ P(X, \phi)]~,
\ee
where $X=1/2 g^{\mu\nu}\partial_{\mu}\phi\partial_{\nu}\phi$ is the kinetic term. We have considered a general Lagrangian $P(X, \phi)$ for the scalar field. The non-minimally coupled function $F(\phi)$ should be positive to guarantee the positiveness of the effective Newton constant. The equations of motions are obtained by the  variation of this action
\bea
& &FG^{\mu\nu}-g^{\mu\nu}\Box F+\nabla^{\mu}\nabla^{\nu}F=- T^{\mu\nu}~,\nonumber\\
& &\nabla_{\mu}(P_{X}\nabla^{\mu}\phi)=P_{\phi}+\frac{1}{2}R F_{\phi}~,
\eea
where $T^{\mu\nu}=-Pg^{\mu\nu}+P_X \nabla^{\mu}\phi\nabla^{\nu}\phi$, $P_X$ represents $\partial P/\partial X$, and $P_{\phi}$ and $F_{\phi}$ are defined in the same way. At the background $ds^2=a^2 (d\eta^2-\delta_{ij}dx^idx^j)$, these equations become
\bea\label{background}
& &\mathcal{H}^2+\mathcal{H}\frac{F'}{F}=\frac{a^2}{3 F}\rho~,\nonumber\\
& &\rho'+3\mathcal{H}(\rho+P)=\frac{3}{a^2}(\mathcal{H}'+\mathcal{H}^2)F'~,
\eea
here we used the reduced Hubble parameter $\mathcal{H}=a'/a$ and defined $\rho=-P+2XP_X$ as usual. There are some requirements for the model building as far as the background equations are concerned.
To solve the flatness and the horizon problems, the absolute value $|\mathcal{H}|=a|H|$ should increase with time, i.e., $d|\mathcal{H}|/d\eta>0$. Furthermore, during the contracting phase $\mathcal{H}<0$,
the energy density $\rho$ is required to increase faster than $a^{-6}$ to suppress the classical anisotropies.

In order to discuss the perturbations and their quantizations we will use the ADM decomposition
$ds^2=N^2 d\eta^2-h_{ij}(dx^i+N^id\eta)(dx^j+N^jd\eta)$ with the lapse function $N$, the shift vector $N^i$ and the induced metric $h_{ij}$.
Following Maldacena \cite{Maldacena:2002vr}, we choose the gauge $\delta\phi=0, h_{ij}=a^2(e^{2\zeta}\delta_{ij}+\gamma_{ij})$. This choice simplifies the calculations and at the same time the left dynamical fields $\zeta$ and $\gamma_{ij}$ are gauge-invariant. The purpose is to find the quadratic action of the scalar perturbation $\zeta$ and the tensor perturbation $\gamma_{ij}$. With the ADM decomposition the action (\ref{action}) is written as
\bea\label{action2}
S &=&\int d^4x \sqrt{h} N\{-\frac{1}{2}F [^{(3)}R+\frac{E^2-E_{ij}E^{ij}}{N^2}\nonumber\\
&-&\frac{2}{N\sqrt{h}}\partial_0(\frac{\sqrt{h}E}{N})
+\frac{2}{N\sqrt{h}}\partial_i(\sqrt{h}E\frac{N^i}{N}+\sqrt{h}h^{ij}\partial_j N)]\nonumber\\
&+&P(\frac{\phi'^2}{2N^2}, \phi)\}~,
\eea
where $h={\rm det}|h_{ij}|$, $E_{ij}=-\frac{1}{2}(h_{ij}'-N_{i|j}-N_{j|i})$ and $E=-h^{ij}E_{ij}$. The extrinsic curvature is
$K_{ij}=E_{ij}/N$.
In these formulae the indices are lowered and raised by the induced metric $h_{ij}$ and its inverse $h^{ij}$, and $N_{i|j}$ represents the covariant derivative of $N_i$ induced by $h_{ij}$.

In linear perturbation theory the scalar and tensor perturbations can be considered separately.
We first consider the scalar perturbation, in this case $h_{ij}=a^2e^{2\zeta}\delta_{ij}$. We first solve for the constraints $N$ and $N^i$ through their equations got from the variation of the action (\ref{action2}) and then plug the result back to (\ref{action2}). At the background level, it is easy to find that $N=a$ and $N^i=0$. When inhomogeneous perturbations are included we need only calculate $N$ and $N^i$ up to the linear order as argued in Ref. \cite{Maldacena:2002vr}. Finally we may get the quadratic action of the scalar perturbation, which has been obtained in \cite{Qiu:2010dk} (see also \cite{Kubota:2011re}). Using our notations the quadratic action is
\be\label{action4}
S^s={1\over 2}\int d^4x (\frac{a\phi'}{\theta'})^2 (\rho_X+\frac{3F_{\phi}^2}{2F})[\zeta'^2-c_s^2(\partial_i\zeta)^2]~,
\ee
where $\theta'=\mathcal{H}+F'/(2F)$, and the square of the sound speed is defined as
\be
c_s^2=\frac{P_X+3F_{\phi}^2/(2F)}{\rho_X+3F_{\phi}^2/(2F)}~.
\ee
We require $\rho_X+3F_{\phi}^2/(2F)>0$ to prevent the $\zeta$ field from being a ghost and $c_s^2>0$ to guarantee the spatial stability. This is different from the minimal coupling case where $\rho_X, P_X>0$ are required. Furthermore, in order to obtain nearly scale-invariant scalar spectrum the factor $(a\phi'/\theta')^2 [\rho_X+3F_{\phi}^2/(2F)]$ should scale approximately as $1/\eta^2$.

Then we focus on the tensor perturbation, for which $h_{ij}=a^2 (e^{\gamma})_{ij}$ and $N=a,~N^i=0$, where $e^{\gamma}$ is the exponential function of the traceless and transverse matrix $\gamma$.
The determinant $h={\rm det}|a^2e^{\gamma}|=a^6 \exp{\rm Tr\gamma}=a^6$ is not perturbed. One can also prove that $E=-h^{ij}E_{ij}=3\mathcal{H}$
is unperturbed up to the second order. With these considerations, we find that the quadratic action for the tensor perturbation is
\be\label{action3}
S^T=\frac{1}{8}\int d^4x a^2F [(\gamma_{ij}')^2-\partial_l\gamma_{ij}\partial_l\gamma_{ij}]~.
\ee
Scale-invariance requires $a^2F\sim 1/\eta^2$.

We use a toy model to illustrate these points. The action is
\be\label{model}
S=\int d^4x \sqrt{g} [\frac{1}{2} \xi^2\phi^2 R-{1\over 2}\partial_{\mu}\phi\partial^{\mu}\phi-V_0(\xi \phi)^q]~,
\ee
so in our notation $F=\xi^2\phi^2$ and $V=V_0 (\xi\phi)^q$. The kinetic term has a wrong sign, this represents a ghost in general relativity and will cause quantum instability. But in our case from above arguments the non-minimal coupling will make the time derivative term of the fluctuation has the right sign if $\rho_X+3F_{\phi}^2/(2F)= 6\xi^2-1\equiv \alpha>0$.
The spatial stability $P_X+3F_{\phi}^2/(2F)>0$ puts the same constraint and $c_s^2=1$. What we will quantize are the fluctuations, so this condition is enough to make this model free from instabilities.
Through definitions of the dimensionless parameters $x\equiv \phi'/(\mathcal{H}\phi),~y\equiv a\sqrt{2V}/(\mathcal{H}\phi)$, the Friedmann equation, i.e., the first equation of background Eqs. (\ref{background}),
may be rewritten as $y^2=x^2+2(1+\alpha)x+1+\alpha $. The second equation of Eqs. (\ref{background}) is
\bea\label{eom}
\dot x=-\frac{y^2}{2\alpha(1+\alpha)}[(6\alpha-\alpha\beta-\beta)x-\beta(1+\alpha)]~,
\eea
where $\dot x\equiv dx/d\ln a$ and we have defined $\beta=4-q$.
This equation has three fixed points corresponding to three scaling solutions:
$x_0=-(1+\alpha)\pm \sqrt{\alpha(1+\alpha)}$ and $x_0=\beta(1+\alpha)/(6\alpha-\alpha\beta-\beta)$.
Both the first and second critical points, in which $y_0=0$, demand $(a\phi'/\theta') \propto \eta^{1/2}$ instead of $1/(-\eta)$ and the produced scalar perturbation has a strong blue tilt. This conflicts with the observations and we will not consider these two points any more.

The third critical point corresponds to the scaling solution
\be\label{solution}
a\propto (-\eta)^p~,~\mathcal{H}=p/\eta~,~
{\rm with}~p=\frac{2(6\alpha-\alpha\beta-\beta)}{\beta^2(1+\alpha)-12\alpha}~.
\ee
One can find that for this scaling solution both pre-factors in the quadratic actions (\ref{action4}) and (\ref{action3}) scale as:
\be\label{scale}
\frac{a\phi'}{\theta'},~a\sqrt{F} \propto (-\eta)^{-1-b}~,
\ee
with $b\equiv \beta^2(1+\alpha)/[(12\alpha-\beta^2(1+\alpha)]$.
Now we study the implications of this solution to the expanding and contracting universes separately. For the expanding universe in which $p<0$ and $\mathcal{H}>0$, the solution (\ref{solution}) can only be stable if $\beta^2(1+\alpha)>36\alpha, ~\beta (1+\alpha)>6\alpha$ or $\beta^2(1+\alpha)<12\alpha, ~\beta(1+\alpha)<6\alpha$. The first case requires $b<-1$ and consequently the generated spectra in such an expanding universe deviate scale-invariance significantly. The second case can achieve scale-invariance if $\beta^2(1+\alpha)\ll 12\alpha$, this will recover the inflation or the slow expansion studied in Refs. \cite{Piao,Qiu} according to the specific parameter space, and we will not consider it any more in this paper.

Now let's consider the contracting universe for which $p>0$ and $\mathcal{H}<0$. Combining with the stability conditions of Eq. (\ref{eom}), the solution (\ref{solution}) is an attractor one if and only if
$\beta^2(1+\alpha)>36\alpha >6\beta (1+\alpha)$ or $\beta^2(1+\alpha)<12\alpha<2\beta(1+\alpha)$. Similarly the first case gives $b<-1$ and leads to non-scale-invariant spectra. We will focus on the second case in which
\be\label{for}
\frac{6\alpha}{1+\alpha}<\beta < \sqrt{\frac{12\alpha}{1+\alpha}}~.
\ee
This requires $0<\alpha<1/2$ and so both $\alpha$ and $\beta$ are small positive parameters.
One can also check that $d|\mathcal{H}|/d\eta=p/\eta^2>0$ in such a contracting phase and this provides the solutions to the flatness and horizon problems.
Furthermore due to the definition of $x$, at the critical point $\phi\propto a^{x_0}$, the energy density $\rho =-\phi'^2/(2a^2)+V$ scales as $a^{x_0(4-\beta)}$. To suppress the classical anisotropies in the contracting universe, $\rho$ must
increase faster than $a^{-6}$, this means $x_0(\beta-4)>6$. Using the expression of $x_0$ this inequality becomes $\beta^2+2\beta<36\alpha/(1+\alpha)$. This condition is not always satisfied in the region (\ref{for}), but for the case
$\beta\ll \sqrt{12\alpha/(1+\alpha)}$ it can be well satisfied.

The scale-invariant scalar and tensor perturbations can be obtained if $\beta\ll \sqrt{12\alpha/(1+\alpha)}$.
In terms of the standard procedure learned from inflation theory one has the spectra $P_s=A_s(k_{\ast}) (k/k_{\ast})^{-2b}$ and $P_t=A_t(k_{\ast}) (k/k_{\ast})^{-2b}$, both have the same small red tilts because $b>3\alpha/(1-2\alpha)$ from
the inequalities (\ref{for}).
The amplitudes $A_s(k_{\ast})$ and $A_t(k_{\ast})$ at the pivot scale $k_{\ast}$ mainly depend on the model parameter $V_0$ which defines the energy scale at which the primordial perturbations were created.
One can calculate that the observational result $A_s(k_{\ast})\sim 10^{-9}$ requires $V_0\sim 10^{-8}$. The tensor-to-scalar ratio is fixed in this model
$r=16b/(1+b)$ and if we choose the parameters $\alpha=0.004,~\beta=0.024$ one can easily find that $r=0.19$ and $n_s-1=n_t\simeq 0.0244$.

So we have seen that with the model (\ref{model})  the nearly scale-invariant scalar and tensor perturbations consistent with the current observations can be obtained in the contracting universe if the parameters $\alpha$ and $\beta$ are positive and small. In terms of them, the action (\ref{model}) is rewritten as
\be\label{conformal}
S=\int d^4x \sqrt{g} [\frac{1+\alpha}{12}\phi^2 R -{1\over  2}\partial_{\mu}\phi\partial^{\mu}\phi-V_0(\sqrt{\frac{1+\alpha}{6}}\phi)^{4-\beta}]~,
\ee
and this shows that non-zero $\alpha$ and $\beta$ represent the breaking of the conformal symmetry. The model in this form was also considered in \cite{Piao}. It is approximately conformal invariant.
We may think that the scale invariances of the spectra are originated from the conformal invariance of the model.

\begin{figure}
\includegraphics[scale=0.4]{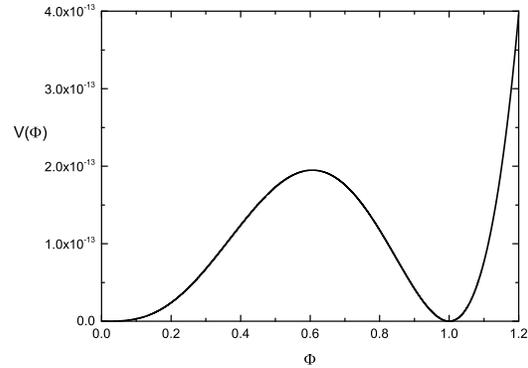}
\caption{The deformed potential, where the dimensionless $\Phi$ is positive definite and $V_0=10^{-8}$, $\beta=0.024$.}
\label{Fig:potential}
\end{figure}

\begin{figure}
\includegraphics[scale=0.4]{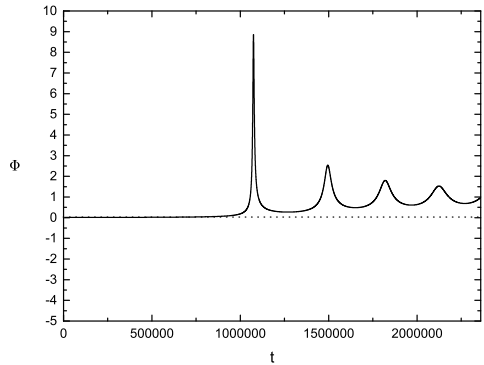}
\caption{ The evolution of $\Phi$. Parameters are the same with those in Fig. \ref{Fig:potential}. We used the cosmic time $t$ instead of the conformal time $\eta$.}
\label{Fig:Phi}
\end{figure}

\begin{figure}
\includegraphics[scale=0.4]{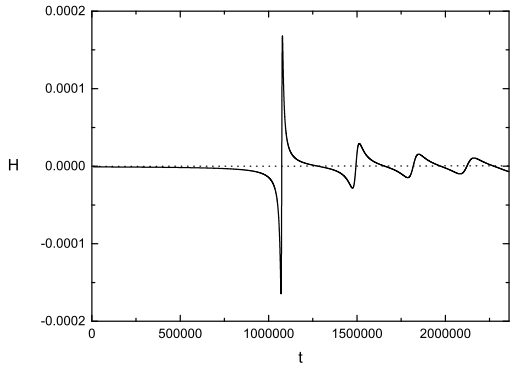}
\caption{The evolution the Hubble parameter.}
\label{Fig:H}
\end{figure}

\begin{figure}
\includegraphics[scale=0.4]{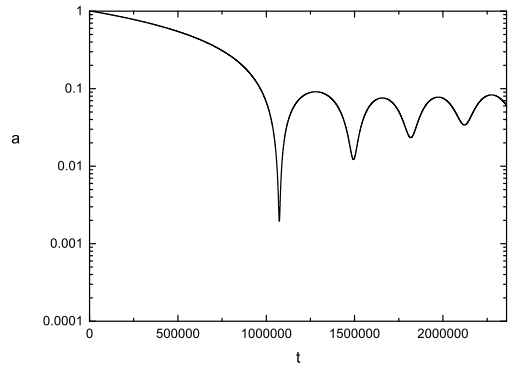}
\caption{The evolution of the scale factor.}
\label{Fig:a}
\end{figure}

It deserves pointing out that this model is not complete. The contracting phase should end at some later time and bounce to an expanding spacetime. This toy model itself does not provide the mechanism of bouncing. For this purpose
we make a little deformation to the toy model (\ref{model}) as follows
\be\label{model2}
V=V_0 \Phi^4 (\Phi^{-\beta}+\Phi^{\beta}-2)~,
\ee
where $\Phi\equiv \xi\phi$. We have added two terms with higher powers to the potential so that it has the form depicted in Fig. \ref{Fig:potential}.
The deformation produces a bump and an extra local minimum in the potential.
The evolution begins at $\Phi\sim 0$. When $\Phi\ll 1$ this deformed model is almost the same with the model (\ref{model}). At this regime $\Phi$ changes slowly and the universe is contracting,
nearly scale-invariant primordial perturbations are generated.
At later time when $\Phi$ is not so small the terms with higher powers become important and the universe bounces to the expansion and then the field $\Phi$
oscillates around the minimum $\Phi=1$ with damped amplitude. Reheating takes place at this stage and the energy of the scalar field is transferred to its produced components such as the radiation. Reheating will make the amplitude of the oscillations decaying more quickly and finally the scalar field itself is frozen at the minimum
$\Phi=1$. The evolution of $\Phi$ with respect to the cosmic time $t$ is plotted in Fig. \ref{Fig:Phi}.
With this frozen value the non-minimal coupling term in the action becomes
\be
\frac{1+\alpha}{12}\phi^2R=\frac{1}{2}\Phi^2 R=\frac{1}{2}R~,
\ee
so after reheating the theory of gravity is identical to general relativity. We also plotted the time evolutions of the Hubble parameter $H=\mathcal{H}/a$
and the scale factor, see Fig. \ref{Fig:H} and Fig. \ref{Fig:a}. One can find that when neglecting reheating the universe would experience multi cycles of expansion and contraction as showed by the damped oscillating behavior of $H$ in Fig. \ref{Fig:H}. Each cycle includes non-singular bounces between expanding and contracting phases, as showed in Fig. \ref{Fig:a}. Gradually the universe approaches to a static universe. This is similar to inflation. Without reheating the inflaton field will oscillates with damped amplitude around the minimum of its potential and the universe enters into the static phase in the future if the cosmological constant vanishes. However, in a real universe 
reheating is necessary and it will interrupt the oscillations and lead the universe to the radiation dominated epoch. In the model considered here, the exit from oscillation to the hot expansion depends on the details of reheating, but we can still get the general idea about how this exit appears. The Friedmann equation (\ref{background}) can be rewritten as \footnote{There are two branches of solutions to the Friedmann equation, besides Eq. (\ref{branch}), another one is $H=-\frac{\dot\Phi}{\Phi}-\sqrt{\frac{\dot\Phi^2}{\Phi^2}+\frac{\rho}{3\Phi^2}}$. But with this branch no bounce will take place.}
\be\label{branch}
H=-\frac{\dot\Phi}{\Phi}+\sqrt{\frac{\dot\Phi^2}{\Phi^2}+\frac{\rho}{3\Phi^2}}~,
\ee
where $\rho$ is the sum of $\rho_{\phi}=-\dot\Phi^2/(2\xi^2)+V$ and the density of radiation $\rho_r$, and we have neglect the contribution from the direct interaction between $\phi$ and radiation. At the early time of oscillation, $\rho_r$ can be neglected, bounces happen at the point $\rho_{\phi}=0$. The density $\rho_{\phi}$ itself also oscillates around the zero point with damped amplitude and reheating speeds up the damping. At the same time, $\rho_r$ increases with time. When it becomes significant, $H$ in Eq. (\ref{branch}) becomes positive definitely, the oscillation stops and  the universe enters into the smooth expanding phase. Reheating finishes at the moment of $\Phi=1$ and $\rho_{\phi}=0$, so that Eq. (\ref{branch}) becomes $H=\sqrt{\rho_r/3}$. This is the familiar Friedmann equation of the hot expanding universe in general relativity. 

Hence we see that sizable gravitational waves suggested by BICEP2 can also be generated in a pre-big bang phase different from inflation.
The price we take is modifying gravity. Using the scalar-tensor theory we showed here that nearly scale-invariant and significant tensor perturbation can be obtained in a contracting universe.  Such tensor perturbation can also be obtained during a slow expanding phase under the same context as pointed out in Refs. \cite{Piao,Qiu} .
It is well known that a scalar-tensor system has different forms in different frames. The frame we discussed above
is usually called Jordan frame and distinguished from the Einstein frame discussed below. It is assumed to be the frame in which
the matter couples to the metric minimally. So that there is no extra force mediated by the scalar field $\phi$ among the matter and this is consistent with
current experiments testing the equivalence principle in the matter sector. But the gravity itself did not obey the strong equivalence principle because the scalar field would mediate a fifth force in the gravity sector. However, in our model (\ref{model2}) this fifth force is not detectable by current gravitational probes because
after bouncing to the hot expansion the scalar field had been stabilized to the minimum $\Phi=1$ through oscillations and decays and the theory of gravity approaches to the general relativity from the beginning of radiation dominated epoch.
In other words, in our model the deviation of the gravity from the general relativity is only significant at the early universe, it does not change the post-big bang history and cannot have effects on later time gravitational probes.

It is also necessary to comment on the possibility of nonsingular bouncing behavior realized in our model. In the general relativity, nonsingular bounce requires
violation of the null energy condition by the matter field. Especially the equation of state of the matter should cross $-1$, similar to the behavior of the quintom dark energy \cite{quintom} at late time. This is not true for the scalar-tensor theory. In fact in the context of scalar-tensor theories effective phantom or quintom dark energy models without violation of null energy condition have been discussed extensively in the literature, see e.g., \cite{starobinsky}. Hence applying the scalar-tensor theory to the early universe, it is also possible to realize a nonsingular bouncing universe without introducing ghosts. Such an example was provided in Ref. \cite{barrow} where a nonsingular universe was obtained in the generalized Brans-Dicke theory without potential. In our model with the quadratic non-minimal coupling and the potential (\ref{model2}), the bounce happened at the point where $\rho=-\phi'^2/(2a^2)+V=0$, but neither the scalar field nor the graviton is ghost.
According to \cite{abramo, starobinsky}, the scalar tensor theory of the type
\be
S=\int d^4x \sqrt{g} [\frac{1}{2} F(\phi) R+{1\over 2}h(\phi)\partial_{\mu}\phi\partial^{\mu}\phi-V(\phi)]~,
\ee
is stable if
\be
F>0,~F_1\equiv Fh+\frac{3}{2}(\frac{dF}{d\phi})^2>0~.
\ee
The first requirement guarantees a positive Newton ``constant" and prevents the graviton from being a ghost, and the second requirement
protects the scalar field from being a ghost. These requirements are fully satisfied in our case.
The non-minimal coupling function $F=\xi^2\phi^2=\Phi^2$ is positive everywhere because $\phi\neq 0$,
and $F_1=-\xi^2\phi^2+6\xi^4\phi^2=\alpha \xi^2\phi^2>0$ because $\alpha$ is a positive parameter as we discussed before.
So in our model, the bounce is stable.

For comparison with the discussions in the Jordan frame, 
it is useful to see what happened in the Einstein frame.
For the toy model (\ref{model}), if we re-scale the metric $\bar{g}_{\mu\nu}=\Omega^2 g_{\mu\nu}$ with
$\Omega=\xi\phi=\Phi$ and through field redefinitions, one may get the action in the Einstein frame,
\be\label{Einstein}
S=\int d^4x\sqrt{\bar{g}}[\frac{1}{2}\bar{R}+\frac{1}{2}\bar{g}^{\mu\nu}\partial_{\mu}\bar{\phi}\partial_{\nu}\bar{\phi}-V_0\exp(-\beta\sqrt{\frac{1+\alpha}{6\alpha}}\bar{\phi})]~.
\ee
The conformal transformation is essentially identical to the redefinitions of the scalar and tensor fields.
The action (\ref{Einstein}) has been used to model the power law inflation in the literature, and the inflation is an attractor solution if $0<\beta\sqrt{(1+\alpha)/(6\alpha)}< \sqrt{2}$.
With the same parameters $\alpha=0.004,~\beta=0.024$, this inflaton has the equation of state $w=-0.992$.
Similarly this inflation model is not complete because it needs other mechanisms to end the inflation.
The deformed model with the potential (\ref{model2}) in the Einstein frame has the potential
$\bar{V}=2V_0 [\cosh(-\beta\sqrt{(1+\alpha)/6\alpha}\bar{\phi}) -1]$ and has the minimum at $\bar{\phi}=0$. With this potential the inflation has a graceful exit.
In the Einstein frame, other matters should couple to the metric non-minimally. However, after inflation, the scalar field $\bar\phi$ has been relaxed to the vacuum $\bar\phi_0=0$ and these non-minimal couplings which depend on the exponential of $\bar\phi$ reduce to the minimal couplings. This means that the Jordan and Einstein frames are identical at late time in our model and denotes again the difference between these two frames are significant only in the early universe.
Though the scalar field is stabilized at the vacuum $\bar\phi_0=0$, its fluctuation still transmits residual force between matter. It is 
straightforwardly to show that the fluctuation around the vacuum $\bar{\varphi}=\bar\phi-\bar\phi_0$ has the potential 
$\bar{V}=(1/2) m_{eff}^2 \bar{\varphi}^2$ with $m_{eff}=\beta\sqrt{V_0(1+\alpha)/3\alpha}$. Please note that we use the unit $M_p=1$ in this paper. With the parameters to produce the right primordial perturbations, the effective mass is $m_{eff}\sim 10^{-5}M_p\sim 10^{13}{\rm GeV}$. Due to this high mass, the residual force is short range and invisible to the fifth force searches. Current experiments show that no 
deviations from the Newton's inverse square law have been found above the distances of $10^{-8}m$ \cite{Adelberger:2003zx,Will}, this places  a lower limit on the effective mass of the scalar field $m_{eff}>10~ {\rm eV}$. In our case the effective mass is well above this limit.

One can show that in both frames the created scalar and tensor perturbations are the same. This reflects the fact that the gauge-invariant scalar and tensor perturbations are frame independent as demonstrated
in Refs. \cite{Prokopec:2013zya,Kubota:2011re}.
Note that the frame or conformal invariance of the scalar perturbation $\zeta$ has relative limited meaning compared with the invariance of gravitational waves.
At least this can be seen from the discussions of \cite{Prokopec:2013zya,Kubota:2011re}. The curvature perturbation $\zeta$ is only invariant under those conformal transformations
$\bar{g}_{\mu\nu}=\Omega^2 g_{\mu\nu}$ in which $\Omega$ is a function of the scalar field $\phi$. However the tensor perturbation which relates to the Weyl tensor is invariant for any $\Omega$.

The conformal invariances of the perturbations have important implications. In terms of the trick from \cite{Bars:2013yba}, the conformal transformations can be upgraded to the gauge transformations.
In fact any scalar-tensor system with canonically normalized kinetic term can be described by the following conformal invariant action,
\be\label{invariant}
S=\int d^4x \sqrt{g} [\frac{\chi^2-\varphi^2}{12}R+\frac{1}{2}(\partial\varphi)^2-\frac{1}{2}(\partial\chi)^2-f(\frac{\varphi}{\chi})(\chi^2-\varphi^2)^2]~,
\ee
where $f$ is an arbitrary function of $\varphi/\chi$. One can prove that this action is invariant under the rescalings
$\bar{g}_{\mu\nu}=\Omega^2(x)g_{\mu\nu},~\bar{\chi}=\Omega^{-1}\chi,~\bar{\varphi}=\Omega^{-1}\varphi $.
Though there introduced two scalar fields, one of them can be gauged away.
The toy model (\ref{conformal}) or (\ref{Einstein}) corresponds to
\be\label{function}
f(\varphi/\chi)=\frac{ V_0}{36}(\frac{\chi+\varphi}{\chi-\varphi})^{-\beta\sqrt{\frac{1+\alpha}{4\alpha}}}~.
\ee
Different frames correspond to different gauges and once the frame is chosen the conformal symmetry is spontaneously broken.
In our example (\ref{function}), if we choose the gauge
\bea
& &\chi=\sqrt{1+\alpha}\phi\cosh[\sqrt{\frac{\alpha}{1+\alpha}}\ln (\sqrt{\frac{1+\alpha}{6}}\phi)]~,\nonumber\\
& &\varphi=\sqrt{1+\alpha}\phi\sinh[\sqrt{\frac{\alpha}{1+\alpha}}\ln (\sqrt{\frac{1+\alpha}{6}}\phi)]~,
\eea
the invariant action (\ref{invariant}) reduces to the action (\ref{conformal}) in the Jordan frame.
As same if we choose the gauge
\be
\chi=\sqrt{6}\cosh\frac{\bar\phi}{\sqrt{6}}~,~\varphi=\sqrt{6}\sinh\frac{\bar\phi}{\sqrt{6}}~,
\ee
it reduces to the action (\ref{Einstein}) in the Einstein frame.
From the discussions of this paper, in the first gauge, the universe was contracting in the pre-big bang era, however in the second gauge the universe was inflating.
Both produce identical nearly scale-invariant perturbations.
From this point, only the conformal invariant quantities, such as the scalar and tensor perturbations are physical because the background evolutions of the early universe, inflation or contraction or other possibilities, depend on the frames and in terms of late time cosmological probes we cannot determine which frame our universe was in.
Perhaps it is more important to pursue the model (\ref{function}) in the conformal invariant action than studying the background evolution in a specific frame.

In summary, we used a model to show that nearly scale-invariant scalar and tensor perturbations can be achieved in the contracting universe in the context of scalar-tensor theory.
This model with slight deformations can bounce to the hot big bang universe and at almost the same time the theory of gravity changes to general relativity and will not change the post-big bang history.
We also note that this model is dual to inflation if we change the frame by the conformal transformation. Different frames correspond to different gauge choices. We want to through this study to reinforce the point that
the background evolution of the early universe is gauge-dependent and has no physical meaning because for the late time observers, we cannot know which
frame our universe was in. What we can measure are the gauge invariant quantities such as the scalar and tensor perturbations.
Hence, if the detection of sizable primordial gravitational waves by BICEP2 is confirmed, we still
cannot say that inflation, the quasi exponentially expansion, must have happened.

{\bf Acknowledgement:} The author is grateful to Yun-Song Piao, Robert Brandenberger for comments on the manuscript and to Youping Wan and Siyu Li for helps in the numerical calculations.
He also thanks Angelika Fertig, Jean-Luc Lehners and Enno Mallwitz for bringing the author's attention to an error of the numerical calculation in the old version. 
This work is supported by Program for New Century Excellent Talents in University and by NSFC under Grants No. 11075074.

{}


\begin{thebibliography}{}

\bibitem{inflation}
  A.~H.~Guth,
  %``The Inflationary Universe: A Possible Solution to the Horizon and Flatness Problems,''
  Phys.\ Rev.\ D {\bf 23} (1981) 347; A.~D.~Linde,
  %``A New Inflationary Universe Scenario: A Possible Solution of the Horizon, Flatness, Homogeneity, Isotropy and Primordial Monopole Problems,''
  Phys.\ Lett.\ B {\bf 108} (1982) 389;
  A.~Albrecht and P.~J.~Steinhardt,
  %``Cosmology for Grand Unified Theories with Radiatively Induced Symmetry Breaking,''
  Phys.\ Rev.\ Lett.\  {\bf 48} (1982) 1220.
  %%CITATION = PRLTA,48,1220;%%

\bibitem{Ade:2013ktc}
  P.~A.~R.~Ade {\it et al.}  [Planck Collaboration],
  %``Planck 2013 results. I. Overview of products and scientific results,''
  arXiv:1303.5062; arXiv:1303.5076;
  arXiv:1303.5082;
  arXiv:1303.5084.


\bibitem{Ade:2014xna}
  P.~A.~R.~Ade {\it et al.}  [BICEP2 Collaboration],
  %``BICEP2 I: Detection Of B-mode Polarization at Degree Angular Scales,''
  arXiv:1403.3985.

\bibitem{Ekpyrotic1}
J.~Khoury, B.~A.~Ovrut, P.~J.~Steinhardt and N.~Turok,
  %``The Ekpyrotic universe: Colliding branes and the origin of the hot big bang,''
  Phys.\ Rev.\ D {\bf 64} (2001) 123522;
P.~J.~Steinhardt and N.~Turok,
  %``Cosmic evolution in a cyclic universe,''
  Phys.\ Rev.\ D {\bf 65} (2002) 126003.

\bibitem{Novello:2008ra}
  M.~Novello and S.~E.~P.~Bergliaffa,
  %``Bouncing Cosmologies,''
  Phys.\ Rept.\  {\bf 463}, 127 (2008).

\bibitem{Wands:1998yp}
  D.~Wands,
  %``Duality invariance of cosmological perturbation spectra,''
  Phys.\ Rev.\ D {\bf 60}, 023507 (1999); F.~Finelli and R.~Brandenberger,
  %``On the generation of a scale invariant spectrum of adiabatic fluctuations in cosmological models with a contracting phase,''
  Phys.\ Rev.\ D {\bf 65}, 103522 (2002).

\bibitem{Cai:2013kja}
  Y.~-F.~Cai, E.~McDonough, F.~Duplessis and R.~H.~Brandenberger,
  %``Two Field Matter Bounce Cosmology,''
  JCAP {\bf 1310}, 024 (2013);
  J.~-Q.~Xia, Y.~-F.~Cai, H.~Li and X.~Zhang,
  %``Evidence for bouncing evolution before inflation after BICEP2,''
  arXiv:1403.7623.

\bibitem{Cheng:2014bma}
 R.~H.~Brandenberger, A.~Nayeri and S.~P.~Patil,
  %``Closed String Thermodynamics and a Blue Tensor Spectrum,''
  arXiv:1403.4927;
  C.~Cheng and Q.~-G.~Huang,
  %``The Tilt of Primordial Gravitational Waves Spectra from BICEP2,''
  arXiv:1403.5463;
H.~Li, J.~-Q.~Xia and X.~Zhang,
  %``Global fitting analysis on cosmological models after BICEP2,''
  arXiv:1404.0238;
  F.~Wu, Y.~Li, Y.~Lu and X.~Chen,
  %``Cosmological parameter fittings with the BICEP2 data,''
  arXiv:1403.6462;
  B.~Hu, J.~-W.~Hu, Z.~-K.~Guo and R.~-G.~Cai,
  %``Reconstruction of the primordial power spectra with Planck and BICEP2,''
  arXiv:1404.3690;
  Y.~Wang and W.~Xue,
  %``Inflation and Alternatives with Blue Tensor Spectra,''
  arXiv:1403.5817;
  M.~Gerbino, A.~Marchini, L.~Pagano, L.~Salvati, E.~Di Valentino and A.~Melchiorri,
  %``Blue Gravity Waves from BICEP2 ?,''
  arXiv:1403.5732;
  A.~Ashoorioon, K.~Dimopoulos, M.~M.~Sheikh-Jabbari and G.~Shiu,
  %``Non-Bunch-Davis Initial State Reconciles Chaotic Models with BICEP and Planck,''
  arXiv:1403.6099.

\bibitem{Piao}
Y.~S.~Piao, arXiv:1109.4266;
arXiv:1112.3737.

\bibitem{Qiu}
T.~Qiu, JCAP {\bf 1206}, 041 (2012).

\bibitem{Clifton:2011jh}
  T.~Clifton, P.~G.~Ferreira, A.~Padilla and C.~Skordis,
  %``Modified Gravity and Cosmology,''
  Phys.\ Rept.\  {\bf 513}, 1 (2012).

\bibitem{Maldacena:2002vr}
  J.~M.~Maldacena,
  %``Non-Gaussian features of primordial fluctuations in single field inflationary models,''
  JHEP {\bf 0305}, 013 (2003).

\bibitem{Qiu:2010dk}
  T.~Qiu and K.~-C.~Yang,
  %``Non-Gaussianities of Single Field Inflation with Non-minimal Coupling,''
  Phys.\ Rev.\ D {\bf 83}, 084022 (2011).

\bibitem{Kubota:2011re}
  T.~Kubota, N.~Misumi, W.~Naylor and N.~Okuda,
  %``The Conformal Transformation in General Single Field Inflation with Non-Minimal Coupling,''
  JCAP {\bf 1202}, 034 (2012).

\bibitem{quintom}
  B.~Feng, X.~-L.~Wang and X.~-M.~Zhang,
  %``Dark energy constraints from the cosmic age and supernova,''
  Phys.\ Lett.\ B {\bf 607}, 35 (2005);
M.~Li, B.~Feng and X.~-m.~Zhang,
  %``A Single scalar field model of dark energy with equation of state crossing -1,''
  JCAP {\bf 0512}, 002 (2005);
  Y.~-F.~Cai, T.~Qiu, Y.~-S.~Piao, M.~Li and X.~Zhang,
  %``Bouncing universe with quintom matter,''
  JHEP {\bf 0710}, 071 (2007).

\bibitem{starobinsky}
  R.~Gannouji, D.~Polarski, A.~Ranquet and A.~A.~Starobinsky,
  %``Scalar-Tensor Models of Normal and Phantom Dark Energy,''
  JCAP {\bf 0609}, 016 (2006).

\bibitem{barrow}
  J.~D.~Barrow,
  %``Nonsingular scalar - tensor cosmologies,''
  Phys.\ Rev.\ D {\bf 48}, 3592 (1993).

\bibitem{abramo}
  L.~R.~Abramo, L.~Brenig, E.~Gunzig and A.~Saa,
  %``On the singularities of gravity in the presence of nonminimally coupled scalar fields,''
  Phys.\ Rev.\ D {\bf 67}, 027301 (2003).

\bibitem{Adelberger:2003zx}
  E.~G.~Adelberger, B.~R.~Heckel and A.~E.~Nelson,
  %``Tests of the gravitational inverse square law,''
  Ann.\ Rev.\ Nucl.\ Part.\ Sci.\  {\bf 53}, 77 (2003).
  
\bibitem{Will}
  C.~M.~Will,
  %``The Confrontation between General Relativity and Experiment,''
  Living Rev.\ Rel.\  {\bf 17}, 4 (2014).

\bibitem{Prokopec:2013zya}
J.~-O.~Gong, J.~-c.~Hwang, W.~-I.~Park, M.~Sasaki and Y.~-S.~Song,
  %``Conformal invariance of curvature perturbation,''
  JCAP {\bf 1109}, 023 (2011);
   T.~Chiba and M.~Yamaguchi,
  %``Extended Slow-Roll Conditions and Rapid-Roll Conditions,''
  JCAP {\bf 0810}, 021 (2008);
  T.~Prokopec and J.~Weenink,
  %``Frame independent cosmological perturbations,''
  JCAP {\bf 1309}, 027 (2013).

\bibitem{Bars:2013yba}
  I.~Bars, S.~-H.~Chen, P.~J.~Steinhardt and N.~Turok,
  %``Antigravity and the Big Crunch/Big Bang Transition,''
  Phys.\ Lett.\ B {\bf 715}, 278 (2012);
     R.~Kallosh and A.~Linde,
  %``Universality Class in Conformal Inflation,''
  JCAP {\bf 1307}, 002 (2013);
I.~Bars, P.~Steinhardt and N.~Turok,
  %``Local Conformal Symmetry in Physics and Cosmology,''
  Phys.\ Rev.\ D {\bf 89}, 043515 (2014);
\end{thebibliography}
\end{document}